\documentclass[twocolumn,prl,superscriptaddress,amssymb,amsmath,floatfix]{revtex4-2}
\usepackage{graphicx}
\usepackage{epstopdf}
\usepackage{comment}
\usepackage{bm}
\usepackage{wrapfig}
\usepackage{times}
\usepackage[colorlinks=true,linkcolor=blue,urlcolor=blue,citecolor=blue]{hyperref}
\usepackage{float}
\usepackage{xcolor}
\usepackage[normalem]{ulem}
\usepackage{flushend}

\newcommand\code[1]{{\tt #1}}
\usepackage{pdfpages} % include pdfs
\usepackage{pgffor} % for loops
\usepackage{xr} % referencing the supplement

\makeatletter
\AtBeginDocument{\let\LS@rot\@undefined}
\makeatother

\def\supplementfilename{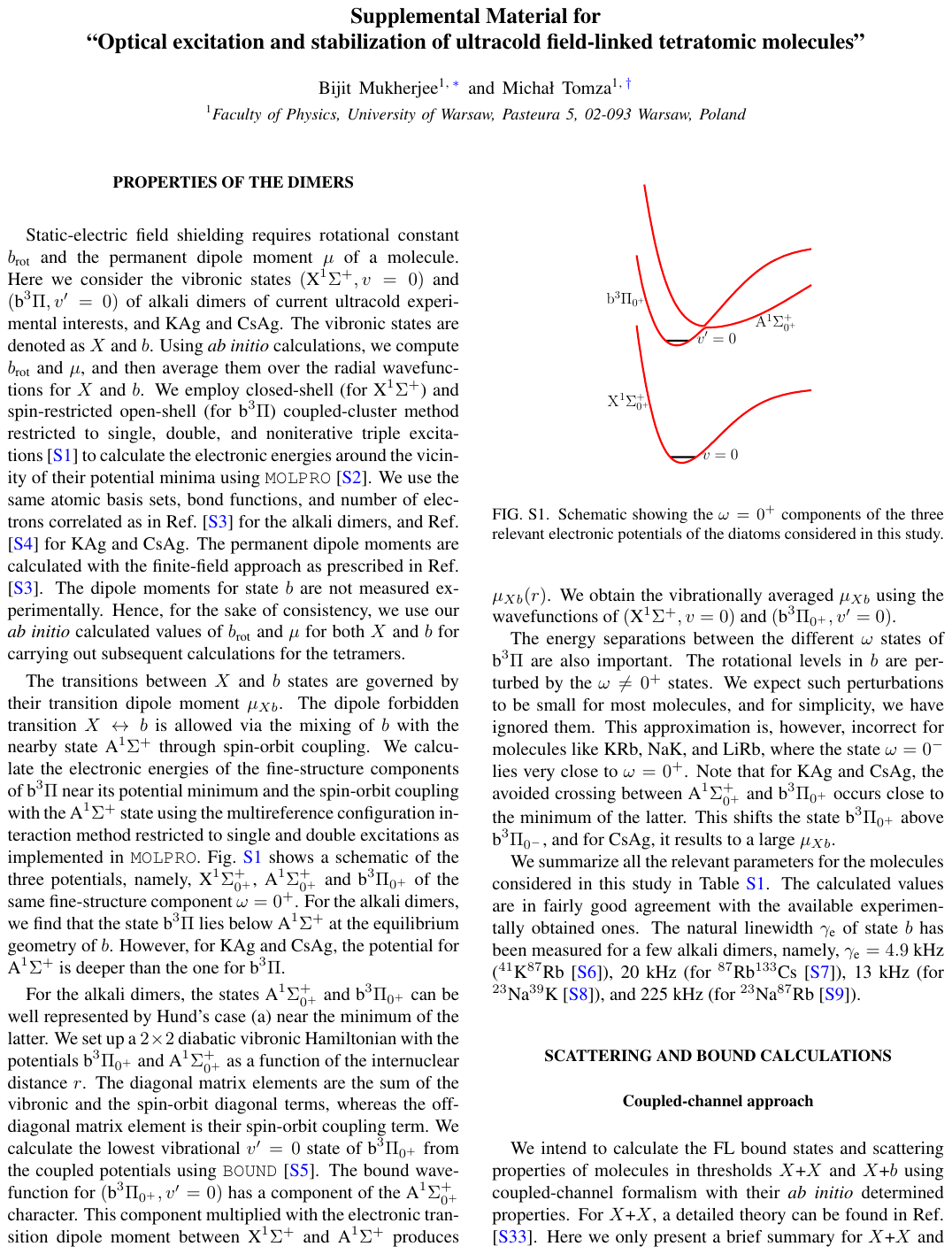}

\pdfximage{\supplementfilename}
\def\numbersupplementpages{\the\pdflastximagepages}

\newif\ifarXiv
\arXivtrue 
\begin{document}
\title{Optical excitation and stabilization of ultracold field-linked tetratomic molecules}

\author{Bijit Mukherjee}
\email{bijit.mukherjee@fuw.edu.pl}
\affiliation{Faculty of Physics, University of Warsaw, Pasteura 5, 02-093 Warsaw, Poland}
\author{Micha{\l} Tomza}
\email{michal.tomza@fuw.edu.pl}
\affiliation{Faculty of Physics, University of Warsaw, Pasteura 5, 02-093 Warsaw, Poland}

\date{\today}

\begin{abstract}
We propose a coherent optical population transfer of weakly bound field-linked (FL) tetratomic molecules (tetramers) to deeper FL bound states using stimulated Raman adiabatic passage. We consider static-electric-field shielded polar alkali-metal diatomic molecules and corresponding FL tetramers in their $\textrm{X}^1\Sigma^+$+$\textrm{X}^1\Sigma^+$ ground electronic state. We show that the excited metastable $\textrm{X}^1\Sigma^+$+$\textrm{b}^3\Pi$ electronic manifold supports FL tetramers in a broader range of electric fields with collisional shielding extended to zero field. We calculate the Franck-Condon factors between the ground and excited FL tetramers and show that they are highly tunable with the electric field. We also predict photoassociation of ground-state shielded molecules to the excited FL states in free-bound optical transitions. We propose proof-of-principle experiments to implement stimulated Raman adiabatic passage and photoassociation using FL tetramers, paving the way for the formation of deeply bound ultracold polyatomic molecules.
\end{abstract}

\maketitle

\textit{Introduction}-- Shielding of polar molecules against collisional loss has been instrumental in the realization of stable ultracold molecular gases. Using external static-electric~\cite{AvdeenkovPRA06, WangNJP15, Gonzalez-MartinezPRA17, MukherjeePRR23}, near-resonant microwave~\cite{KarmanPRL18, LassablierePRL18, KarmanPRA19, KarmanPRXQ25} or laser~\cite{XiePRL20} fields, long-range repulsion between a pair of polar molecules can be engineered by controlling their dipole-dipole interaction. Static-field and microwave shielding have been successfully implemented~\cite{MatsudaScience20, LiNP21, AndereggScience21, BigagliNP23, LinPRX23}, where the latter paved the way for achieving Fermi degeneracy~\cite{SchindewolfNature22} and Bose-Einstein condensation of dipolar molecules~\cite{BigagliNature24}.

The external field responsible for generating the shielding repulsion at long range also creates an attractive dipolar interaction at longer range. The resulting potential well can be deep enough to support one or more quasibound states, also known as ``field-linked'' (FL) states. Existence of FL states was predicted 20 years ago~\cite{AvdeenkovPRL03}, and they have been studied for static-electric- and microwave-field scenarios~\cite{HuangPRA12, QuemenerPRL23, MukherjeePRR24}. Only recently, they have been observed in the collisions of microwave-shielded Na$^{40}$K molecules~\cite{ChenNature23} as scattering resonances. Using such resonances, weakly bound ultracold tetratomic molecules (hereafter ``tetramers'') have been associated in the long range from pairs of diatomic Na$^{40}$K molecules~\cite{ChenNature24}. 

At the same time, there is a growing interest in creating ultracold polyatomic molecules. They provide unique opportunities in studies of cold chemistry~\cite{BalakrishnanJCP16, KarmanNP24}, precision measurements~\cite{HutzlerQST20, DoyleJPSCP22}, exotic quantum phases~\cite{SchmidtPRR22}, and quantum information processing~\cite{AlbertPRX20}. A few triatomic species have been successfully laser cooled~\cite{PrehnPRL16, KozyryevPRL17, MitraScience20, VilasNature22, LasnerPRL25}; however, they need favorable rovibronic structure to establish closed optical cycles. This limits the number of polyatomic molecules that can be cooled directly to ultracold temperatures. FL tetramers, on the other hand, may provide alternate pathways for obtaining ground-state stable ultracold polyatomic molecules~\cite{ChenNature24}.

An extremely successful method for creating ultracold diatomic molecules (hereafter ``dimers'') from pairs of ultracold atoms is magnetoassociation followed by stimulated Raman adiabatic passage (STIRAP)~\cite{ChinRMP2010, BergmannRMP98}. STIRAP involves two steps where the population from an initial weakly bound state $|i\rangle$ is optically transferred to a target deeply bound state $|f\rangle$ via an excited intermediate state $|e\rangle$. The $|i\rangle \rightarrow |e\rangle$ is the excitation transition, whereas the $|e\rangle \rightarrow |f\rangle$ is the stabilization step. Here, we ask whether it is possible to realize a STIRAP transfer of weakly bound FL tetramer states ($|i\rangle$) at the long range to deeper bound states ($|f\rangle$) at the short range, and thus extend the tool from dimers to tetramers. There are however twofold challenges. First, to locate an isolated $|e\rangle$ state from a very high density of states for the tetramers. Second, to determine the full sequence of STIRAP transfer of the FL tetramers to their absolute ground state, which requires a full knowledge of their interaction potential and transition dipole moment surfaces.

\begin{figure}[tbh]
 \includegraphics[width=\columnwidth]{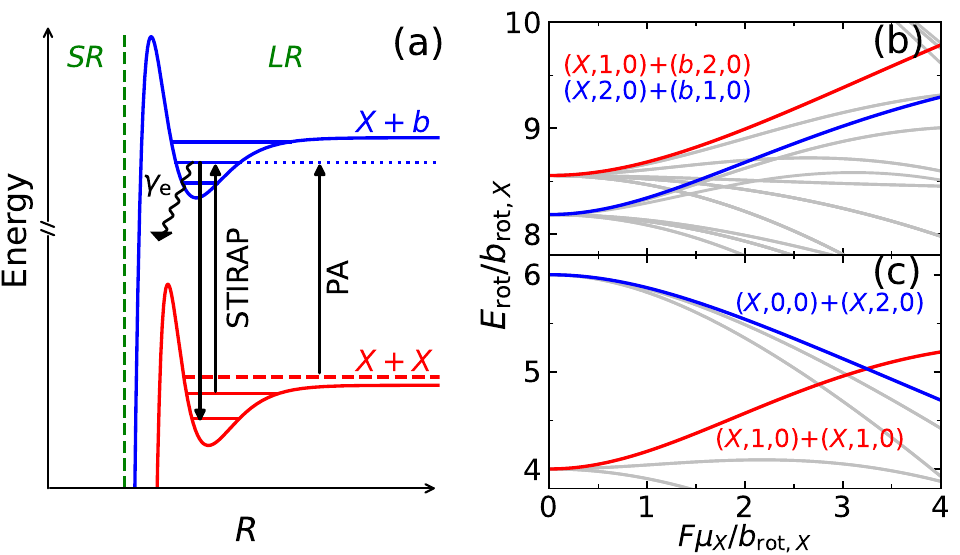}
\caption{(a) Schematic showing the envisaged STIRAP using FL tetramers (solid horizontal lines). Molecules with more than one FL state in $X$+$X$ may be used for demonstrating STIRAP via FL states of $X$+$b$. Additionally, the PA of shielded dimers colliding in $X$+$X$ (red dashed line) to FL states in $X$+$b$ may be envisaged. The natural linewidth $\gamma_\textrm{e}$ of the $b$ state determines the lifetime of the FL states in $X$+$b$. A representative vertical green dashed line separates the short (SR) and long (LR) ranges. Panels (b) and (c): Energies $E_\textrm{rot}$ of the rotor pair levels of LiCs for thresholds (b) $X$+$b$ and (c) $X$+$X$ as a function of $F$. Panel (c) is universal for any polar diatomic molecule in state $X$. However, energies in panel (b) depend on the differences in $\mu$ and $b_\textrm{rot}$ for states $X$ and $b$, so they are molecule-dependent. }%
     \label{fig:schematic}
 \end{figure}

In this Letter, we focus on the first challenge and show that FL states exist in a vibronically excited manifold that can serve as $|e\rangle$. Such states have the advantage that they are well isolated and can be controlled via the external field. To this end, we consider FL tetramers formed from shielded polar alkali-metal dimers (and similar molecules) colliding in their ground vibronic state $(\textrm{X}^1\Sigma^+, v=0)$ (hereafter ``$X$''). We use an external static-electric field $\boldsymbol{F}$ for shielding and show that FL tetramers coexist in the ground $X$+$X$ and the excited electronic manifold $X$+$b$ [Fig.\ \ref{fig:schematic}(a)], where $b$ denotes the long-lived excited vibronic state $(\textrm{b}^3\Pi_{0^+}, v'=0)$. We choose such excited FL tetramer states as $|e\rangle$, and study Franck-Condon factors (FCFs) between the FL states of $X$+$X$ and $X$+$b$ as functions of $F=|\boldsymbol{F}|$, with NaCs and LiCs as representative examples.

%In this work, we consider FL tetramers formed from shielded polar alkali-metal dimers (and similar molecules) colliding in their ground vibronic state $(\textrm{X}^1\Sigma^+, v=0)$ (hereafter ``$X$''). To realize STIRAP, we propose an intermediate state $|e\rangle$ embedded in their excited electronic manifold $X$+$b$, where $b$ represents the long-lived excited vibronic state $(\textrm{b}^3\Pi_{0^+}, v'=0)$ of an alkali dimer. We use an external static-electric field $\boldsymbol{F}$ for shielding of molecules colliding in $X$+$X$, and show that FL tetramers coexist in both $X$+$X$ and $X$+$b$ manifolds, as shown schematically in Fig.\ \ref{fig:schematic}(a). We choose such \revision{vibronically} excited FL tetramer states as our $|e\rangle$ states, and study bound-bound transitions between the FL states of $X$+$X$ and $X$+$b$ as functions of $F=|\boldsymbol{F}|$, with NaCs and LiCs as representative examples. Such $|e\rangle$ states have the advantage that they are well isolated, and can be controlled via the external field. 

Modeling the stabilization step to the absolute ground state of tetramers remains a challenge. However, in the following, we show that FL states of $X$+$b$ are potential candidates for achieving this goal. For this, we propose proof-of-principle experiments for implementing STIRAP to transfer population from a shallower FL state in $X$+$X$ to a deeper one within the same long-range potential well via an excited FL state of $X$+$b$ [Fig.\ \ref{fig:schematic}(a)]. We demonstrate this with LiCs, which supports two FL states in $X$+$X$ in the presence of $F$. Additionally, we study photoassociation (PA) prospects for converting shielded dimers of $X$+$X$ to FL tetramers in $X$+$b$. We consider RbCs and LiCs to demonstrate PA to FL states in the excited $X$+$b$ manifold. We choose bosonic molecules, but our treatment is equally applicable to the fermionic ones. %\revision{In the following, we discuss the rationale behind choosing $b$ as the excited vibronic state of the dimer.}

\textit{The metastable $b$ state}-- The electronic spin-forbidden transitions between $\textrm{X}^1\Sigma^+_{0^+}$ and $\textrm{b}^3\Pi_{0^+}$ of an alkali dimer are weakly allowed only through the spin-orbit mixing of the latter with $\textrm{A}^1\Sigma^+_{0^+}$. The subscript $0^+$ is the value of the projection $\omega$ of the total electronic angular momentum along the dimer axis. Owing to a large FCF between the vibronic states $b$ and $X$, the optical transition $X \leftrightarrow b$ is observable, but has a very narrow linewidth. The radiative decay of $b$ to $\textrm{a}^3\Sigma^+$ manifold is also very slow, making $b$ a metastable state. The natural linewidth $\gamma_e$ of $b$ for a few alkali dimers has been measured~\cite{KobayashiPRA14, BausePRL20, HeNJP21, GregoryNP24}, and except for NaRb, $\gamma_e$ is on the order of tens of kHz (see Supplemental Material (SM) at \cite{sup-mat-X_b}). As shown below, the fact that $\gamma_\textrm{e}$ is much smaller than the binding energies of the FL states in $X$+$b$ enables their optical addressing and manipulating. Moreover, for an imperfect STIRAP transfer, it is an advantage to have a metastable intermediate state to mitigate any population loss.

\textit{Molecules in $X$ and $b$ states}-- In the presence of a static-electric field $\boldsymbol{F}$, the effective Hamiltonian for a spin-free diatomic molecule A in vibronic state $X$ is 
\begin{equation}\label{eq:molham}
\hat{h}_\textrm{A} = \hat{h}_\textrm{rot,A} + \hat{h}_\textrm{Stark,A}.
\end{equation}
Here $\hat{h}_{\textrm{rot,A}}=b_{\textrm{rot,A}} \hat{\boldsymbol{j}}^2$ is the rotational Hamiltonian, where $b_{\textrm{rot,A}}$ is the rotational constant and $\hat{\boldsymbol{j}}$ is the molecular rotation. $\hat{h}_{\textrm{Stark,A}}=-\boldsymbol{\mu}_\textrm{A} \cdot \boldsymbol{F}$, where $\boldsymbol{\mu}_\textrm{A}$ is the dipole moment along the dimer axis. Alkali dimers have nuclear spins which give rise to an additional hyperfine Hamiltonian $\hat{h}_\textrm{hf}$~\cite{BrownBook03}. However, the effects of $\hat{h}_\textrm{hf}$ on their scattering properties under effective shielding are very small~\cite{MukherjeeNJP25, MukherjeePRR25}; hence we ignore them.

Molecules in $\textrm{b}^3\Pi$ have additional terms in their effective Hamiltonian in Eq.\ (\ref{eq:molham}). The dominant one is the spin-orbit coupling Hamiltonian $\hat{h}_\textrm{so}$, which splits a $^3\Pi$ term into the fine-structure components $^3\Pi_{\omega}$, with $|\omega|=0^{\pm},1,2$. The $\pm$ of $\omega=0$ represents the symmetry under reflection through a plane containing the molecular axis. Around the minimum of the potential, the $|\omega|>0$ levels are typically separated from $\omega=0^{\pm}$ by tens of cm$^{-1}\times hc$~\cite{sup-mat-X_b}, where $h$ and $c$ are, respectively, the Planck constant and the speed of light. The degeneracy of $\omega=0^{\pm}$ is lifted due to the interactions with other electronic states of the same symmetry. Their energy gap is typically tens of GHz$\times h$ for most molecules~\cite{sup-mat-X_b}. The remaining terms of $\textrm{b}^3\Pi$ Hamiltonian are namely, the orbital and spin contributions in the rotational Hamiltonian denoted by $\hat{h}_\textrm{rso}$, the spin-spin $\hat{h}_\textrm{ss}$, the spin-rotation $\hat{h}_\textrm{sr}$, the $\Lambda$-doubling $\hat{h}_\textrm{LD}$, and hyperfine $\hat{h}_\textrm{hf}$ terms. We denote these additional terms collectively as $\hat{h}_\textrm{add} = \hat{h}_\textrm{rso} +\hat{h}_\textrm{ss} +\hat{h}_\textrm{sr} +\hat{h}_\textrm{LD} +\hat{h}_\textrm{hf}$. In the Hund's case (a) basis, $\hat{h}_\textrm{add}$ has both diagonal and off-diagonal matrix elements in $\omega$~\cite{BrownBook03}. 

Our methodology involves only the lowest few rotational levels of the vibronic state $b$. Since $|\omega|>0$ states are energetically far away compared to the rotational spacings, $\hat{h}_\textrm{add}$ contributes small first- and second-order perturbative corrections to the rotational energy levels in $b$. To a first approximation, we ignore these additional contributions as they do not affect our methodology qualitatively. Moreover, we ignore the terms in $\hat{h}_\textrm{Stark}$ that are off-diagonal in $\omega=0^+$and $0^-$. This is again a justified approximation for most molecules where the energy gap between $\omega=0^+$ and $0^-$ components is large compared to $b_\textrm{rot}$, but, it is not appropriate for KRb, NaK, and LiRb. Therefore, we consider the same effective Hamiltonian $\hat{h}$, Eq.\ (\ref{eq:molham}), for both $X$ and $b$ molecules, which simplifies our treatment enormously. We will use the subscripts $X$ and $b$ for $b_\textrm{rot}$ and $\mu$ to denote their values obtained after averaging over the respective vibronic states. We calculate these parameters using \textit{ab initio} methods~\cite{LadjimiPRA24} as implemented in \code{MOLPRO}~\cite{MOLPRO_brief} for alkali dimers and similar molecules of current experimental interest~\cite{sup-mat-X_b}. We find that the ratio $b_\textrm{rot}/\mu$ is comparable for $X$ and $b$ molecules, allowing them to be polarizable to a similar extent in the presence of $F$. We label the dimer levels by $(\beta,\tilde{j},m)$, where $\beta$ represents the vibronic state ($X$ or $b$), $\tilde{j}$ correlates with the free-rotor quantum number $j$ at zero field, and $m$ is its conserved projection along $\boldsymbol{F}$.

\begin{figure}[tbh]
\includegraphics[width=0.48\textwidth]{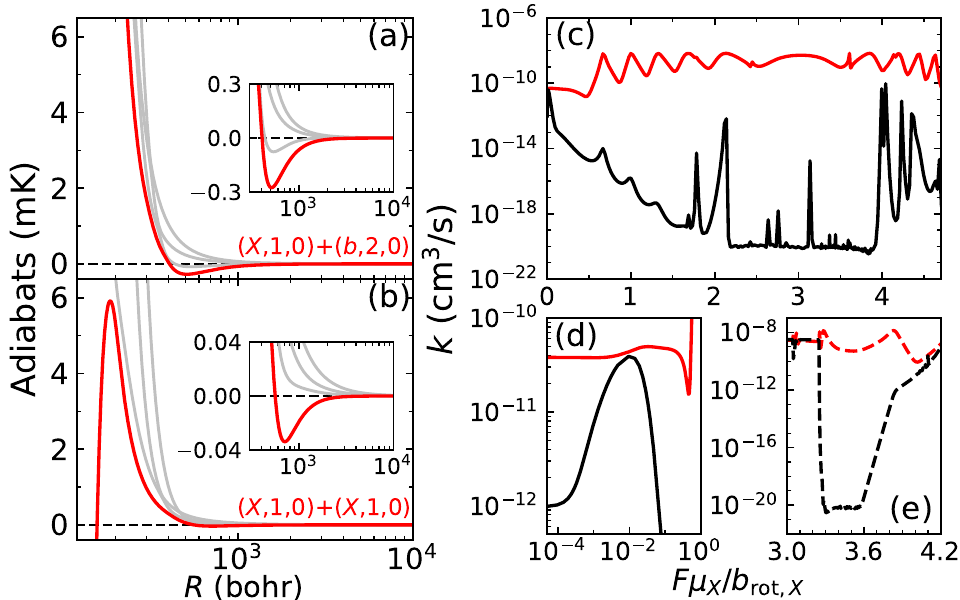}
\caption{Panels (a) and (b): Adiabats for LiCs correlating to partial waves $L=0,2,4,6$ for the initial threshold (a) ($X, 1, 0)$+$(b,2,0)$, and (b) $(X,1,0)$+$(X,1,0)$ at $F=3.57 b_{\textrm{rot},X}/\mu_X$. The adiabats are shown relative to their initial thresholds. The incoming s-wave ($L=0$) channel is shown in red. Insets show expanded views of the long-range potential wells that support FL tetramer states. Panels (c)-(e): Rate coefficients $k$ for elastic scattering (red) and total two-body loss (black) for collisions in $X$+$b$ [solid curves in (c) and (d)] and $X$+$X$ [dashed curves in (e)].}%
     \label{fig:LiCs_adia_ratecoeff}
\end{figure}

\textit{Shielding and FL states in $X$+$X$}-- The interaction between a pair of polar molecules A and B involved in shielding, occurring at long range, is given by their dipole-dipole interaction $\hat{H}_\textrm{dd} = [\boldsymbol{\mu}_\textrm{A} \cdot\boldsymbol{\mu}_\textrm{B} - 3(\boldsymbol{\mu}_\textrm{A}\cdot\hat{\boldsymbol{R}}) (\boldsymbol{\mu}_\textrm{B}\cdot\hat{\boldsymbol{R}})] / (4\pi\epsilon_0 R^3)$, where $R$ is the intermolecular distance and $\hat{\boldsymbol{R}}$ is a unit vector along the intermolecular axis. Shielding may occur when two pair levels that are connected by $\hat{H}_\textrm{dd}$ are close enough in energy such that they are strongly mixed. For static-field shielding, molecules colliding in the state $(\beta,\tilde{j},m) = (X,1,0)$ experience repulsion due to mixing with the lower lying pair level $(X,0,0)$+$(X,2,0)$ at fields above $F=3.244b_{\textrm{rot},X}/\mu_X$ [Fig.\ \ref{fig:schematic}(c)]. The repulsion, however, becomes weak for $F > 3.8b_{\textrm{rot},X}/\mu_X$.

Shielding may be understood by adiabats that are the $R$-dependent eigenvalues of $\hat{h}_\textrm{A} + \hat{h}_\textrm{B} + \hbar^2 \hat{\boldsymbol{L}}^2/(2\mu_\textrm{red}R^2) + \hat{H}_\textrm{dd}$. Here $\hat{\boldsymbol{L}}$ is the angular momentum operator for relative rotation of the two molecules and $\mu_\textrm{red}$ is the corresponding reduced mass. Figure \ref{fig:LiCs_adia_ratecoeff}(b) shows the adiabats for LiCs for different partial waves $L$ correlating with the initial pair level $(X,1,0)$+$(X,1,0)$ at $F=3.57b_{\textrm{rot},X}/\mu_X$ ($7.5$ kV/cm). It demonstrates that the adiabats are repulsive for $R<400$ bohr. In addition, $\hat{H}_\textrm{dd}$ couples the incoming $L=0$ with $L=2$ channel to produce an attraction at larger distances that is asymptotically proportional to $-d^4/R^4$, where $d$ is the induced dipole moment along $\boldsymbol{F}$. The resulting potential well is shown in the inset of Fig.\ \ref{fig:LiCs_adia_ratecoeff}(b). This potential well for molecules with larger values of $\mu_X$ can be deep enough to support FL states. For example, NaRb and NaCs support one FL state each, LiCs supports two, and KAg and CsAg~\cite{SmialkowskiPRA21} support more than two FL states~\cite{MukherjeePRR24}. Here, we perform scattering calculations with \code{MOLSCAT}~\cite{HutsonMolscatCPC19} to determine the rate coefficients for elastic scattering $k_\textrm{el}$ and two-body loss $k_{2,\textrm{loss}}$ for various molecules using their \textit{ab initio} determined $\mu$ and $b_\textrm{rot}$. We solve coupled-channel equations using a fully absorbing boundary condition at the short range~\cite{ClaryFDCS87, JanssenPhD12}. We follow the methodology of Ref.\ \cite{MukherjeePRR23}, but with a slightly different basis set~\cite{sup-mat-X_b}. Figure \ref{fig:LiCs_adia_ratecoeff}(e) shows $k_\textrm{el}$ and $k_{2,\textrm{loss}}$ for LiCs at a collision energy $E_\textrm{coll}=10$ nK$\times k_\textrm{B}$, where $k_\textrm{B}$ is the Boltzmann constant. The peaks in $k_\textrm{el}$ indicate the fields where FL states cross the threshold.

\textit{Shielding and FL states in $X$+$b$ manifold}-- Shielding in $X$+$b$ originates from a different physics. Owing to the difference in $b_\textrm{rot}$ and $\mu$ for $X$ and $b$, any two pair levels $(X,\tilde{j},m)$+$(b,\tilde{j}+1,m)$ and $(X,\tilde{j}+1,m)$+$(b,\tilde{j},m)$ have distinct energies. Two such pair levels for LiCs (with $b_{\textrm{rot},b}>b_{\textrm{rot},X}$) are shown as colored lines in Fig.\ \ref{fig:schematic}(b). This means molecules in the upper level, here $(X,1,0)$+$(b,2,0)$, experience repulsion due to mixing via $\hat{H}_\textrm{dd}$ with the lower lying $(X,2,0)$+$(b,1,0)$. This feature exists in all fields, including $F=0$, which is an important and surprising finding in this study. Note that molecules with $b_{\textrm{rot},X}>b_{\textrm{rot},b}$ will be shielded in $(X,2,0)$+$(b,1,0)$ due to coupling to the lower lying $(X,1,0)$+$(b,2,0)$ level. 

There are two contributions to $\hat{H}_\textrm{dd}$ for molecules interacting in two different vibronic states. The first is between the permanent dipoles in $X$ and $b$, and the other is due to a resonant dipolar interaction between a pair of superposition states of $X$ and $b$. The former is responsible for shielding repulsion in the upper level. The latter is proportional to $\mu_{Xb}^2$, where $\mu_{Xb}$ is the $X \leftrightarrow b$ transition dipole moment, and thus is usually smaller. The details of the interaction potential, the coupled-channel approach, and the basis sets are given in SM~\cite{sup-mat-X_b}. Figure \ref{fig:LiCs_adia_ratecoeff}(a) shows the adiabats for different $L$ correlating to the pair level $(X,1,0)$+$(b,2,0)$ for LiCs at $F=7.5$ kV/cm, a field where shielding in $(X,1,0)$+$(X,1,0)$ is optimal. They are repulsive at $R<350$ bohr. Hereafter, we use $X$+$X$ and $X$+$b$ to denote the thresholds $(X,1,0)$+$(X,1,0)$ and $(X,1,0)$+$(b,2,0)$, respectively. The inset in Fig.\ \ref{fig:LiCs_adia_ratecoeff}(a) shows the long-range potential well for $X$+$b$, which is deeper than the one in $X$+$X$ (since $\mu_b>\mu_X$). The well can support up to six FL states. Subsequently, we determine $k_\textrm{el}$ and $k_{2,\textrm{loss}}$ for LiCs for a broad range of fields at $E_\textrm{coll}=10$ nK$\times k_\textrm{B}$, as shown in Fig.\ \ref{fig:LiCs_adia_ratecoeff}(c) and (d). It is evident that $k_{2,\textrm{loss}}$ for $X$+$b$ collisions is highly suppressed at most fields, including $F=0$ and the ones relevant for shielding in $X$+$X$. A discussion on the behavior of $k_{2,\textrm{loss}}$ is given in SM~\cite{sup-mat-X_b}.

\begin{figure}[tb]
\includegraphics[width=0.48\textwidth]{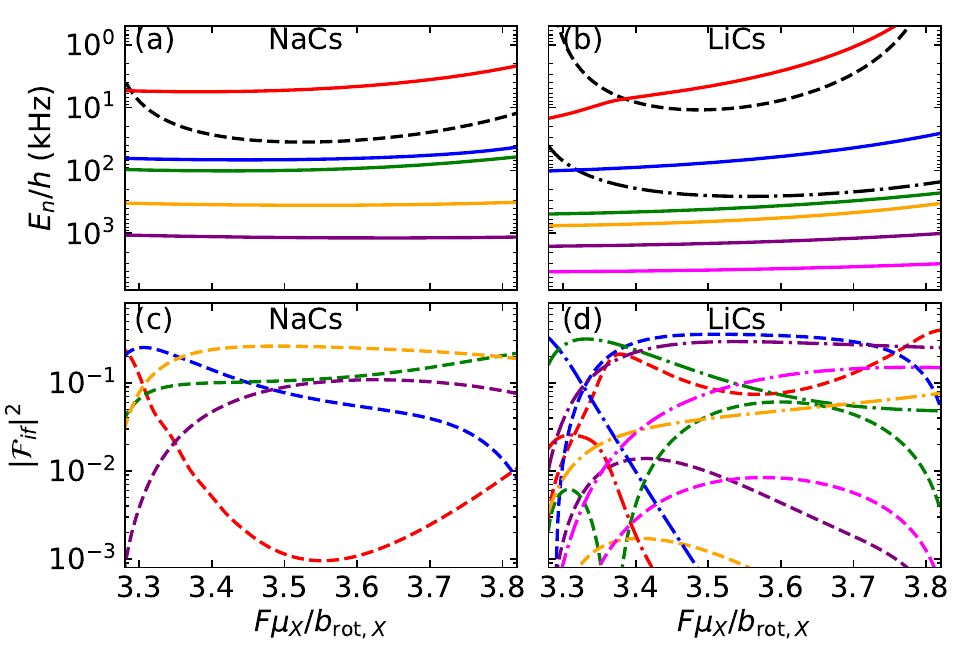}
\caption{Upper panels show binding energies $E_n$ of FL tetramer states for (a) NaCs and (b) LiCs in their $X$+$X$ (dashed black) and $X$+$b$ (solid colored) thresholds as functions of $F$. The FCFs are shown in the lower panels. Their dash type and colors correlate to the tetramer states of $X$+$X$ and $X$+$b$, respectively, from the upper panel.}%
     \label{fig:Tetramer_bound_fcfs}
\end{figure}

\textit{FCFs and STIRAP prospects}-- Here, we study the bound-bound transitions between the FL states of $X$+$X$ and $X$+$b$. We calculate the binding energies $E_n$ of the FL tetramers as functions of $F$ using coupled-channel methods~\cite{MukherjeePRR24} interfaced with \code{BOUND}~\cite{HutsonBoundCPC19}. We choose fields where shielding in both $X$+$X$ and $X$+$b$ is effective. The number of the FL states supported in the $X$+$X$ and $X$+$b$ thresholds for various molecules is given in Table S2 in SM~\cite{sup-mat-X_b}. The binding energies of the FL states of NaCs and LiCs are shown in the upper panels of Fig.\ \ref{fig:Tetramer_bound_fcfs}. Note that the FL states in both $X$+$X$ and $X$+$b$ are quasibound with a finite energy width. However, static-electric-field shielding can achieve extreme two-body loss suppression [see Fig.\ \ref{fig:LiCs_adia_ratecoeff}(c) and (e)], making the energy widths of the FL states very narrow~\cite{sup-mat-X_b}. Thus, for $X$+$b$ FL states, their decay is almost fully determined by the decay of the $b$ molecules.

We calculate the FCFs $(|\mathcal{F}_{if}|^2)$ between the tetramer states $i$ of $X$+$X$ and states $f$ of $X$+$b$ as shown in the lower panels of Fig.\ \ref{fig:Tetramer_bound_fcfs}. The quantities $|\mathcal{F}_{if}|^2$ contain the radial plus the angular overlap of the tetramer wave functions as described in SM~\cite{sup-mat-X_b}. The FCFs are highly tunable with the electric field, particularly for LiCs, which has multiple FL states in the ground $X$+$X$ manifold. Thus, transitions between FL tetramers are controllable with the external field via the tunability of their FCFs. This is unlike the scenario for the dimers, where FCFs cannot be tuned easily~\cite{TomzaMP13}. The transition dipole moments $D_{if}$ between the FL states are calculated from the FCFs using the relation $D_{if} = \mu_{Xb} |\mathcal{F}_{if}|$~\cite{sup-mat-X_b}. The transition wavelength $\lambda_0$ is mostly determined by the vibronic transition energy $T_{Xb}$ between $X$ and $b$. Among the alkali dimers, $\lambda_0$ ranges from 850 to 1200 nm, whereas for KAg and CsAg, $\lambda_0 \sim$ 500 and 550 nm, respectively~\cite{sup-mat-X_b}.

Using our calculated FCFs for the FL tetramers of LiCs, a proof-of-principle STIRAP experiment can be envisaged, in which the population from the shallower tetramer state of $X$+$X$ can be transferred to the deeper one. The above stabilization reduces the vibrationally averaged interdimer distance $\langle R \rangle_v$ from 2800 to 950 bohr, nearly by a factor of 3. Similar proof-of-principle transfers can be realized for KAg and CsAg, which accommodate multiple FL states in $X$+$X$~\cite{MukherjeePRR24}. $\langle R \rangle_v$ for the deepest FL state in $X$+$b$ is about 850 and 600 bohr for NaCs and LiCs, respectively, which are much smaller than the ones in $X$+$X$. This might be promising in generating FCFs with deeper rovibrational states embedded inside the long-range barrier of $X$+$X$. An accurate determination of the intermediate and short-range interactions is needed to verify this. It should be noted that the STIRAP laser intensities should be low enough such that the Rabi couplings and the differential ac Stark shifts are smaller than the level spacings~\cite{sup-mat-X_b}. 

\textit{Photoassociation prospects}-- PA involves absorption of a photon by a pair of colliding atoms to form an excited diatomic molecule~\cite{LettARPC95, JonesRMP06}. There have been theoretical proposals on extending PA to atom-molecule~\cite{PerezRiosPRL15, ElkamshishyJPCA23} and molecule-molecule~\cite{GacesaPRR21} collisions, where the former have been recently observed~\cite{CaoPRL24}.

Here we consider PA prospects by studying free-bound transitions of shielded dimers colliding in $X$+$X$ to FL tetramers in $X$+$b$. We calculate the PA rate coefficient $k_\textrm{PA}$ using the relation $k_\textrm{PA}= (gh/2\mu_\textrm{red}k_0)\sum_{n} |S_{kn}|^2$~\cite{NapolitanoPRL94, JonesRMP06}. Here $g=2$ for identical molecules colliding in $X$+$X$ and $k_0=\sqrt{2\mu_\textrm{red}E_\textrm{coll}/\hbar^2}$. The PA process for the transition of scattering states $|k\rangle$ of $X$+$X$ to bound states $|n\rangle$ of $X$+$b$ is encapsulated by the energy dependent S-matrix element $S_{kn}(k_0)$, defined as
\begin{align}
    |S_{kn}|^2 = \frac{\Gamma^\textrm{s}_n\Gamma^\textrm{d}_n}{(E_\textrm{coll} + hc/\lambda -hc/\lambda_0)^2 + \frac{1}{4}[\Gamma^\textrm{s}_n + \Gamma^\textrm{d}_n]^2}.
\end{align}
Here $\lambda$ is the PA laser wavelength, $hc/\lambda_0$ is the energy of $|n\rangle$ relative to $X$+$X$, and $\Gamma^\textrm{s}_n$ and $\Gamma^\textrm{d}_n$ are, respectively, the energy widths for the stimulated and spontaneous decay of $|n\rangle$. The stimulated width is energy-dependent and is defined as $\Gamma^\textrm{s}_n(k_0) = (4\pi^2I/\epsilon_0 c) D^2_{kn}(k_0)$, where $\epsilon_0$ is the vacuum permittivity, $I$ is the laser intensity, and we approximate $\Gamma^d_n$ by $\gamma_\textrm{e}$ of the $b$ molecules. 

\begin{figure}[tbp]
\includegraphics[width=0.48\textwidth]{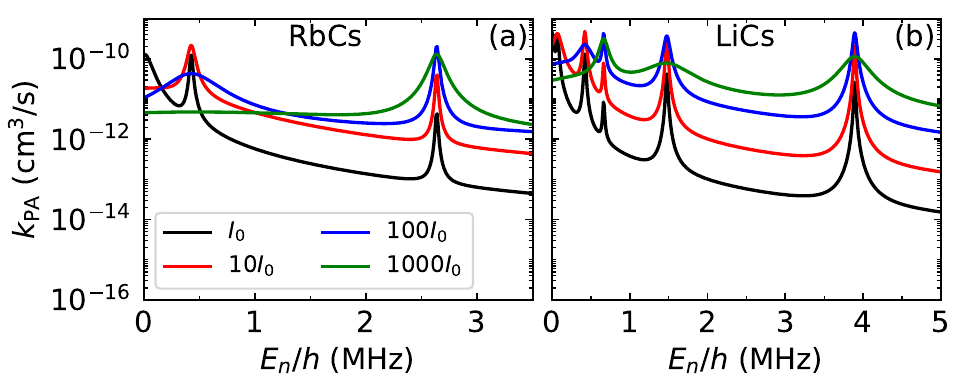}
    \caption{Photoassociation rates $k_\textrm{PA}$ for colliding dimers in $X$+$X$ with $E_\textrm{coll}=500$ nK$\times k_\textrm{B}$ as functions of the binding energy of $X$+$b$ tetramer for (a) RbCs ($F=2.82$ kV/cm, $\lambda_0=1190$ nm) and (b) LiCs ($F=7.30$ kV/cm, $\lambda_0=1160$ nm) for different laser intensities with $I_0=5$ $\mu$W/cm$^2$.}
    \label{fig:PA_spectra}
\end{figure}

Figure \ref{fig:PA_spectra} shows the PA spectra for RbCs+RbCs and LiCs+LiCs at fixed electric fields for different laser intensities. We consider $\gamma_\textrm{e}=20$ kHz$\times h$ for both molecules. We observe well-resolved PA signals for both molecules. However, the intensities required are small. This is mainly due to (1) the requirement that $\Gamma^\textrm{s}_n \ll E_n$, and (2) the long-range nature of the scattering wave functions of $X$+$X$ and bound wave functions of $X$+$b$ produce large values of $|\mathcal{F}_{kn}|^2$, thus requiring smaller $I$. Figure~\ \ref{fig:PA_spectra} shows that $I \gtrsim 5$ mW/cm$^2$ destroys the PA peaks. Also, the PA rates become smaller as $E_\textrm{coll}$ of the colliding dimers in $X$+$X$ increases~\cite{sup-mat-X_b}. For RbCs, FL states do not exist in ground $X$+$X$. Nevertheless, our methods show that the PA of RbCs+RbCs to FL tetramers in the $X$+$b$ manifold is possible. Similar physics is expected for KCs molecules~\cite{sup-mat-X_b}.

\textit{Outlook} -- The FL tetramers in $X$+$b$ provide an excellent tool to optically control ultracold tetramers. Our predictions can be extended to microwave (MW) shielded molecules, which offer higher tunability. One can control the ellipticity of the MW polarization~\cite{ChenNature23} or can implement double MW fields~\cite{KarmanPRXQ25} to make the long-range potential well deeper. This may allow at least two FL states in $X$+$X$ and implement our proposed STIRAP pathway with more molecules. However, this may also increase the two-body loss of the dimers~\footnote{At least two FL states have been realized in MW-shielded Na$^{40}$K molecules by tuning to large ellipticity in MW polarization, however very high losses have been observed in such a scenario~\cite{ChenNature23}.}. Coupled-channel calculations are required to verify this. Since our methodology is based on the long-range dipolar interaction of the dimers, our predictions are not susceptible to experimental or theoretical uncertainties in determinations of the molecular potentials at the short range.

Our ultimate goal is to transfer the FL states to deeply bound tetramers near the global minimum of the ground interaction potential. There might be two possible strategies. First, the transfer to deeply bound tetramer states is envisaged via $X$+$b$ intermediate states. For this, one might lower and shift the repulsive barrier in $X$+$b$ to shorter distances by controlling the external field parameters. This will increase the FCFs with the $X$+$X$ states. Second, the deepest $X$+$X$ FL state, which gets populated through our proposed STIRAP pathway, can be coupled to a deeply bound rovibrational state within $X$+$X$ via a subsequent STIRAP sequence. These directions require knowledge of shorter-range tetramer potentials in both the ground and excited electronic manifolds. We intend to address these in our future studies.

\begin{acknowledgements}
\textit{Acknowledgements}-- We thank Matthew Frye, Xin-Yu Luo, Andreas Schindewolf, and Jun Ye for valuable discussions. We gratefully acknowledge the European Union (ERC, 101042989 -- QuantMol and MSCA, 101203827 -- UltracoldTetramers) for financial support and Poland’s high-performance computing infrastructure PLGrid (HPC Center: ACK Cyfronet AGH) for providing computer facilities and support (computational grant no.~PLG/2024/017844).
\end{acknowledgements}

\textit{Data availability}--The data that support the findings of this article are openly available \cite{MukherjeeXbData2025}.

%\newpage
\bibliography{references}

%-------------------------------------------------------------
% Code to include supp mat at the end
% Source: https://github.com/agdelma/IncludeSupplement
%-------------------------------------------------------------
\ifarXiv
    

\section*{Supplementary material}

# Supplemental Material for
# "Optical excitation and stabilization of ultracold field-linked tetratomic molecules"
Bijit Mukherjee[1, *] and Michał Tomza[1, †]

[1]*Faculty of Physics, University of Warsaw, Pasteura 5, 02-093 Warsaw, Poland*
(Dated: June 19, 2025)

## S1. PROPERTIES OF THE DIMERS

Static-electric field shielding requires rotational constant $b_{\rm rot}$ and the permanent dipole moment $\mu$ of a molecule. Here we consider the vibronic states $(X^1\Sigma^+, v=0)$ and $(b^3\Pi, v'=0)$ of alkali dimers of current ultracold experimental interests, and KAg and CsAg. The vibronic states are denoted as $X$ and $b$. Using *ab initio* calculations, we compute $b_{\rm rot}$ and $\mu$, and then average them over the radial wavefunctions for $X$ and $b$. We employ closed-shell (for $X^1\Sigma^+$) and spin-restricted open-shell (for $b^3\Pi$) coupled-cluster method restricted to single, double, and noniterative triple excitations [S1] to calculate the electronic energies around the vicinity of their potential minima using MOLPRO [S2]. We use the same atomic basis sets, bond functions, and number of electrons correlated as in Ref. [S3] for the alkali dimers, and Ref. [S4] for KAg and CsAg. The permanent dipole moments are calculated with the finite-field approach as prescribed in Ref. [S3]. The dipole moments for state $b$ are not measured experimentally. Hence, for the sake of consistency, we use our *ab initio* calculated values of $b_{\rm rot}$ and $\mu$ for both $X$ and $b$ for carrying out subsequent calculations for the tetramers.

The transitions between $X$ and $b$ states are governed by their transition dipole moment $\mu_{Xb}$. The dipole forbidden transition $X \leftrightarrow b$ is allowed via the mixing of $b$ with the nearby state $A^1\Sigma^+$ through spin-orbit coupling. We calculate the electronic energies of the fine-structure components of $b^3\Pi$ near its potential minimum and the spin-orbit coupling with the $A^1\Sigma^+$ state using the multireference configuration interaction method restricted to single and double excitations as implemented in MOLPRO. Fig. S1 shows a schematic of the three potentials, namely, $X^1\Sigma^+_{0+}$, $A^1\Sigma^+_{0+}$ and $b^3\Pi_{0+}$ of the same fine-structure component $\omega = 0^+$.

For the alkali dimers, the states $A^1\Sigma^+_{0+}$ and $b^3\Pi_{0+}$ can be well represented by Hund's case (a) near the minimum of the latter. We set up a $2\times 2$ diabatic vibronic Hamiltonian with the potentials $b^3\Pi_{0+}$ and $A^1\Sigma^+_{0+}$ as a function of the internuclear distance $r$. The diagonal matrix elements are the sum of the vibronic and the spin-orbit diagonal terms, whereas the off-diagonal matrix element is their spin-orbit coupling term. We calculate the lowest vibrational $v'=0$ state of $b^3\Pi_{0+}$ from the coupled potentials using BOUND [S5]. The bound wavefunction for $(b^3\Pi_{0+}, v'=0)$ has a component of the $A^1\Sigma^+_{0+}$ character. This component multiplied with the electronic transition dipole moment between $X^1\Sigma^+$ and $A^1\Sigma^+$ produces

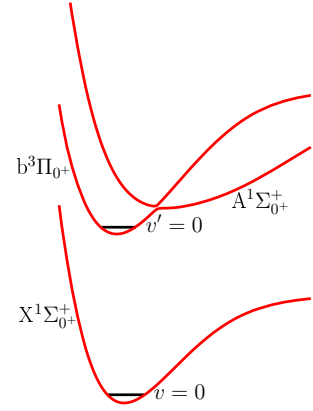

FIG. S1. Schematic showing the $\omega = 0^+$ components of the three relevant electronic potentials of the diatoms considered in this study.

$\mu_{Xb}(r)$. We obtain the vibrationally averaged $\mu_{Xb}$ using the wavefunctions of $(X^1\Sigma^+, v=0)$ and $(b^3\Pi_{0+}, v'=0)$.

The energy separations between the different $\omega$ states of $b^3\Pi$ are also important. The rotational levels in $b$ are perturbed by the $\omega \neq 0^+$ states. We expect such perturbations to be small for most molecules, and for simplicity, we have ignored them. This approximation is, however, incorrect for molecules like KRb, NaK, and LiRb, where the state $\omega = 0^-$ lies very close to $\omega = 0^+$. Note that for KAg and CsAg, the avoided crossing between $A^1\Sigma^+_{0+}$ and $b^3\Pi_{0+}$ occurs close to the minimum of the latter. This shifts the state $b^3\Pi_{0+}$ above $b^3\Pi_{0-}$, and for CsAg, it results to a large $\mu_{Xb}$.

We summarize all the relevant parameters for the molecules considered in this study in Table S1. The calculated values are in fairly good agreement with the available experimentally obtained ones. The natural linewidth $\gamma_e$ of state $b$ has been measured for a few alkali dimers, namely, $\gamma_e = 4.9$ kHz ($^{41}$K$^{87}$Rb [S6]), 20 kHz (for $^{87}$Rb$^{133}$Cs [S7]), 13 kHz (for $^{23}$Na$^{39}$K [S8]), and 225 kHz (for $^{23}$Na$^{87}$Rb [S9]).

## S2. SCATTERING AND BOUND CALCULATIONS

### A. Coupled-channel approach

We intend to calculate the FL bound states and scattering properties of molecules in thresholds $X+X$ and $X+b$ using coupled-channel formalism with their *ab initio* determined properties. For $X+X$, a detailed theory can be found in Ref. [S32]. Here we only present a brief summary for $X+X$ and

---

* bijit.mukherjee@fuw.edu.pl
† michal.tomza@fuw.edu.pl

TABLE S1. Parameters of the $X^1\Sigma^+$ and $b^3\Pi$ potentials for different molecules arranged in increasing order of their dipole moment. $r_e$ is the equilibrium distance in bohr. $D_e$ is the well depth in cm$^{-1}$. $b_{\rm rot}$ is the rotational constant in MHz. $\mu$ and $\mu_{Xb}$ are the permanent and transition dipole moments in debye. The ratio $b_{\rm rot}/\mu$ is in kV/cm. $T_e$ is the $X \to b$ transition energy in cm$^{-1}$. The quantities $\Delta E_{0\pm} = E_{b,0^-} - E_{b,0^+}$ and $\Delta E_{01} = E_{b,1} - E_{b,0^+}$ are respectively the energy differences (in cm$^{-1}$) between the $v' = 0$ levels of the fine-structure components $\omega = 0^-$ and $0^+$, and $\omega = 1$ and $0^+$ of $b^3\Pi$. The experimental values, rounded off to the same significant digits as for the values of the present work, are also shown.

| Molecule | $X^1\Sigma^+$ properties | | | | | $b^3\Pi$ properties | | | | | | | | Refs. |
|---|---|---|---|---|---|---|---|---|---|---|---|---|---|---|
| | $r_e$ | $D_e$ | $b_{\rm rot}$ | $\mu$ | $b_{\rm rot}/\mu$ | $r_e$ | $D_e$ | $T_e$ | $b_{\rm rot}$ | $\mu$ | $\mu_{Xb}$ | $\Delta E_{0\pm}$ | $\Delta E_{01}$ | |
| $^{41}$K$^{87}$Rb | 7.67 | 4150 | 1100 | 0.63 | 3.47 | 7.61 | 7120 | 9610 | 1120 | 1.31 | 0.27 | 1.0 | 53 | This work |
| | 7.69 | 4220 | 1100 | 0.57 | 3.83 | 7.61 | 7080 | 9620 | 1120 | - | - | - | - | [S6, S10, S11] |
| $^{87}$Rb$^{133}$Cs | 8.36 | 3740 | 491 | 1.21 | 0.81 | 8.19 | 6500 | 8600 | 511 | 2.05 | 0.55 | 7.0 | 120 | This work |
| | 8.37 | 3840 | 490 | 1.22 | 0.80 | 8.20 | 6300 | 8720 | 510 | - | 0.68 | - | 93 | [S12–S15] |
| $^{39}$K$^{133}$Cs | 8.08 | 3980 | 915 | 1.86 | 0.98 | 7.91 | 6670 | 8670 | 955 | 3.25 | 0.48 | 3.6 | 91 | This work |
| | 8.10 | 4070 | 912 | - | - | 7.90 | 6600 | 8830 | 960 | - | - | - | 105 | [S16–S18] |
| $^{23}$Na$^{39}$K | 6.61 | 5220 | 2850 | 2.71 | 2.09 | 6.62 | 6750 | 11470 | 2840 | 3.05 | 0.11 | 0.12 | 13 | This work |
| | 6.61 | 5270 | 2848 | 2.72 | 2.08 | 6.61 | 6700 | 11560 | 2850 | - | 0.11 | 0.06 | 15 | [S8, S19–S21] |
| $^{23}$Na$^{87}$Rb | 6.89 | 4970 | 2080 | 3.30 | 1.25 | 6.88 | 6400 | 11230 | 2090 | 4.35 | 0.40 | 1.4 | 42 | This work |
| | 6.88 | 5030 | 2090 | 3.2 | 1.30 | 6.87 | 6380 | 11310 | 2100 | - | 0.40 | - | 50 | [S9, S22–S24] |
| $^7$Li$^{87}$Rb | 6.55 | 5890 | 6420 | 4.00 | 3.19 | 6.40 | 8380 | 10170 | 6790 | 5.60 | 0.17 | 0.014 | 36 | This work |
| | 6.55 | 5920 | 6470 | 4.00 | 3.21 | - | - | - | - | - | - | - | - | [S25, S26] |
| $^{23}$Na$^{133}$Cs | 7.28 | 4870 | 1730 | 4.53 | 0.76 | 7.16 | 6200 | 10030 | 1790 | 6.04 | 0.84 | 7.4 | 89 | This work |
| | 7.28 | 4954 | 1736 | 4.8 | 0.72 | 7.14 | 6081 | 10240 | 1805 | - | - | - | 94 | [S26–S28] |
| $^7$Li$^{133}$Cs | 6.94 | 5810 | 5600 | 5.29 | 2.10 | 6.64 | 8170 | 9000 | 6110 | 6.81 | 0.30 | 4.7 | 78 | This work |
| | 6.93 | 5875 | 5604 | 5.5 | 2.02 | 6.67 | 8063 | 9175 | 6048 | - | 0.5 | - | 76 | [S29–S31] |
| $^{39}$K$^{107}$Ag | 5.61 | 13200 | 2000 | 8.52 | 0.47 | 5.75 | 6110 | 20080 | 1900 | 6.97 | 0.09 | $-1.6$ | 47 | This work |
| $^{133}$Cs$^{107}$Ag | 6.11 | 13560 | 814 | 9.76 | 0.17 | 6.12 | 6480 | 18260 | 810 | 8.85 | 2.24 | $-83$ | 140 | This work |

present the formalism for handling collisions in $X + b$.

In the presence of an external static electric field $F$ along $Z$, the rotational states are field-dressed due to the Stark interaction. They are denoted $|\beta, \tilde{j}, m\rangle$ and correlate at zero field to free-rotor states $|\beta, j, m\rangle$. The quantum number $\beta$ labels the vibronic state $X$ or $b$. The projection $m$ is conserved and

$$|\beta, \tilde{j}, m\rangle = \sum_j c_{j\tilde{j}}^{\beta,m}(F)|\beta, j, m\rangle. \quad (S1)$$

We diagonalize the molecular Hamiltonian $\hat{h}$ as given in Eq. (1) of the main text to calculate the coefficients $c_{j\tilde{j}}^{\beta,m}$ at each field $F$.

For a pair of colliding molecules A and B, the Hamiltonian reads

$$\hat{H} = \frac{\hbar^2}{2\mu_{\rm red}}\left(-R^{-1}\frac{d^2}{dR^2}R + \frac{\hat{\boldsymbol{L}}}{R^2}\right) + \hat{h}_{\rm A} + \hat{h}_{\rm B} + V_{\rm int}, \quad (S2)$$

where $V_{\rm int}$ is the interaction potential.

In the coupled-channel approach, the total wavefunction $\Psi$ is expanded

$$\Psi(R, \hat{\boldsymbol{R}}, \hat{\boldsymbol{r}}_{\rm A}, \hat{\boldsymbol{r}}_{\rm B}) = R^{-1}\sum_i \Phi_i(\hat{\boldsymbol{R}}, \hat{\boldsymbol{r}}_{\rm A}, \hat{\boldsymbol{r}}_{\rm B})\psi_i(R), \quad (S3)$$

where $\psi_i(R)$ are the radial wavefunctions, and $\hat{\boldsymbol{r}}_{\rm A(B)}$ is a unit vector along the axis of molecule A(B). We use a basis set of functions,

$$|\Phi_i\rangle = \hat{P}_\eta|\beta_{\rm A}, \tilde{j}_{\rm A}, m_{\rm A}\rangle|\beta_{\rm B}, \tilde{j}_{\rm B}, m_{\rm B}\rangle|L, M_L\rangle, \quad (S4)$$

where $|L, M_L\rangle$ are the eigenfunctions of $\hat{\boldsymbol{L}}^2$. The operator $\hat{P}_\eta$ is the permutation for identical molecules with $\eta = 1(-1)$ for symmetric (antisymmetric) combination

$$\hat{P}_\eta|{\rm A}\rangle|{\rm B}\rangle|L, M_L\rangle = \frac{1}{\sqrt{2(1+\delta_{\rm AB})}}$$
$$\times \Big(|{\rm A}\rangle|{\rm B}\rangle + \eta|{\rm B}\rangle|{\rm A}\rangle\Big)|L, M_L\rangle, \quad (S5)$$

where $|{\rm A(B)}\rangle \equiv |\beta_{\rm A(B)}, \tilde{j}_{\rm A(B)}, m_{\rm A(B)}\rangle$. The total parity $\epsilon = \eta(-1)^L$ is $+(-)$ for identical bosonic (fermionic) molecules.

### B. Interaction potential

Shielding occurs due to dipole-dipole interactions at intermolecular distances $R \gg 100$ bohr. At such distances, neither the chemical interactions nor the higher multipole terms dominate. We therefore approximate $V_{\rm int} \approx \hat{H}_{\rm dd}$, defined in the main text. For molecules in the same vibronic state as in $X + X$, $\hat{H}_{\rm dd}$ expressed in the spherical tensor form is

$$\hat{H}_{\rm dd} = -\frac{\sqrt{6}}{4\pi\epsilon_0 R^3}T^{(2)}(\hat{\mu}_X, \hat{\mu}_X)C^{(2)}(\hat{\boldsymbol{R}}), \quad (S6)$$

where $T^{(2)}$ and $C^{(2)}$ are second-rank tensors, $\hat{\boldsymbol{R}}$ is the unit vector along the intermolecular axis and the components of $C^2(\hat{\boldsymbol{R}})$ are Racah-normalized spherical harmonics.

For molecules in different vibronic states as in $X + b$, $\hat{H}_{\text{dd}}$ has two contributions

$$\hat{H}_{\text{dd}} = -\frac{\sqrt{6}}{4\pi\epsilon_0 R^3} C^{(2)}(\hat{\boldsymbol{R}}) \left[ T^{(2)}(\hat{\mu}_X, \hat{\mu}_b) + T^{(2)}(\hat{\mu}_{Xb}, \hat{\mu}_{Xb}) \right]. \tag{S7}$$

The first term in the square brackets is the interaction between the permanent dipoles of $X$ and $b$ states. The second is the resonant dipolar interaction between the symmetrized pair states proportional to the $X \leftrightarrow b$ transition dipole moment. Since $\mu_{Xb}^2 < \mu_X \mu_b$, the second contribution is smaller but non-negligible for a few molecules.

### C. Basis sets

We include field-dressed rotor functions up to $\tilde{j}_{\max} = 5$ for each molecule. However, this gives a basis set too large to be used directly in coupled-channel calculations. We therefore divide the basis functions into two groups, namely "class 1" and "class 2", according to the pairs of rotor functions involved. The class 1 pair functions are used explicitly in the coupled-channel calculations, while the class 2 functions are taken into account through Van Vleck transformations as described in Ref. [S32]. We choose rotor pair levels in class 1 that are closest in energy to the initial level, and move the rest to class 2.

For collisions in $X+X$, the initial rotor level is $(\beta, \tilde{j}, m) = (X, 1, 0) + (X, 1, 0)$, so we include functions with $(X, \tilde{j} \leq 2, |m| \leq \min(\tilde{j}, 1))$ in class 1. This produces a total number of rotor pairs $N_{\text{rot}} = 28$. The reason for restricting $|m|$ to 1 is that there is no first-order matrix element in $\hat{H}_{\text{dd}}$ that can connect states with $\Delta m > 1$, and $m$ in the initial rotor levels is 0.

For collisions in $X + b$, a larger number of functions in class 1 are required due to the difference in the $\mu$ and $b_{\text{rot}}$ values. The initial rotor level is $(\beta_{\text{A}}, \tilde{j}_{\text{A}}, m_{\text{A}}) + (\beta_{\text{B}}, \tilde{j}_{\text{B}}, m_{\text{B}}) = (X, 1, 0) + (b, 2, 0)$. We choose a total of $N_{\text{rot}} = 39$ functions in class 1, which are closest to the initial level. With the restrictions $|m|, |m'| \leq 1$, they are: $(X, 1, m) + (b, 1, m')$ (×9); $(X, 0, 0) + (b, 2, m)$ (×3); $(X, 2, m) + (b, 0, 0)$ (×3); $(X, 1, m) + (b, 2, m')$ (×9); $(X, 2, m) + (b, 1, m')$ (×9); $(X, 0, 0) + (b, 3, m)$ (×3); and $(X, 3, m) + (b, 0, 0)$ (×3). The number in the parentheses indicates the number of functions for a given combination.

The transition energy between the rotor pair levels in $X+X$ and $X + b$ is $T = T_e + \Delta E_{\text{rot}}$, where $\Delta E_{\text{rot}}$ is the difference in the rotational energies at a given $F$. Since $T \gg b_{\text{rot}}$, we do not consider functions with $\beta_{\text{A}} = \beta_{\text{B}} = X$ or $b$ in the basis set for calculations in $X + b$.

For the basis set in the partial waves, we consider $L$ up to $L_{\max} = 20$. Since $\hat{H}_{\text{dd}}$ conserves the parity of $L$, this means it is restricted to take only even (odd) values for the bosons (fermions). In the presence of $F$, the projection of the total angular momentum for the colliding pair $M_{\text{tot}} = m_{\text{A}} + m_{\text{B}} + M_L$ is conserved. We perform calculations only for $M_{\text{tot}} = 0$ for which the s-wave incoming channel is included in the basis set. This is a reasonable approximation at a low collision energy $E_{\text{coll}} = 10 \text{ nK} \times k_{\text{B}}$ considered here.

For calculations of the FL bound states, we consider a much smaller set of class 1 functions. This is because the potential well supporting the bound states has a dominant character of the initial rotor pair level. There is only a small admixture of the rotor level that lies just below the initial level, and is responsible for the shielding repulsion in the latter. So, we include only $(X, 1, 0) + (X, 1, 0)$ and $(X, 0, 0) + (X, 2, 0)$ levels for $X + X$ molecules, and $(X, 1, 0) + (b, 2, 0)$ and $(X, 2, 0) + (b, 1, 0)$ levels for $X + b$ molecules in class 1. Also, we restrict $L_{\max}$ to 6.

The above basis sets produce bound and scattering properties converged to within 1%.

### D. Rate coefficients and FL states

A colliding pair of molecules that reaches short range is likely to be lost through processes that may include inelastic transitions, chemical reactions, or laser absorption. To model these processes, we solve the coupled equations subject to a fully absorbing boundary condition at short range [S33, S34] using MOLSCAT [S35]. The numerical methods used are as described in Ref. [S32]. Table S2 summarizes the ratio $\gamma = k_{\text{el}}/k_{2,\text{loss}}$ and the number of FL states for molecules that show effective shielding at field $\tilde{F}_X = 3.5$.

### E. Suppression of two-body loss in $X + b$

The difference in $b_{\text{rot}}$ for states $X$ and $b$ gives rise to non-resonant dipolar interaction between two pair levels $|1\rangle \equiv (X, \tilde{j}, m) + (b, \tilde{j}+1, m)$ and $|2\rangle \equiv (X, \tilde{j}+1, m) + (b, \tilde{j}, m)$. In second-order perturbation theory, the upper lying level experiences repulsion proportional to $|H_{12}|^2/\Delta E_{12}$, where $H_{12}$ is the coupling matrix element and $\Delta E_{12}$ is their energy separation. $H_{12}$ is proportional to the product $\mu_X \mu_b$, and $|\Delta E_{12}|$ scales with $|b_{\text{rot},b} - b_{\text{rot},X}|$. This repulsion achieves shielding against two-body loss when molecules collide in the higher

TABLE S2. Scattering and bound-state properties at field $\tilde{F}_X = F\mu_X/b_{\text{rot},X} = 3.5$. The ratio $\gamma = k_{\text{el}}/k_{2,\text{loss}}$ is shown. #FLT indicates the number of FL states supported by the long-range potential well for the incoming channel. The subscripts g and e represent $X + X$ and $X + b$, respectively.

| Molecule | $\log_{10}(\gamma_g)$ | #FLT$_g$ | $\log_{10}(\gamma_e)$ | #FLT$_e$ |
|---|---|---|---|---|
| $^{87}$Rb$^{133}$Cs | 3.3 | 0 | 1.6 | 3 |
| $^{23}$Na$^{39}$K | 4.0 | 0 | 2.9 | 0 |
| $^{39}$K$^{133}$Cs | 5.0 | 0 | 2.4 | 3 |
| $^{23}$Na$^{87}$Rb | 7.9 | 1 | 8.5 | 2 |
| $^{7}$Li$^{87}$Rb | 9.2 | 1 | 11 | 3 |
| $^{23}$Na$^{133}$Cs | 11 | 1 | 11 | 4 |
| $^{7}$Li$^{133}$Cs | 11 | 2 | 11 | 5 |
| $^{39}$K$^{107}$Ag | 12 | 2 | 10 | 4 |
| $^{133}$Cs$^{107}$Ag | 12 | 4 | 11 | 6 |

lying pair state (between $|1\rangle$ and $|2\rangle$). The repulsion persists from zero field to very high fields, and eventually dies off due to the interaction of the initial level with other thresholds lying above.

The shielding interaction in $X + b$ for LiCs is illustrated in Fig. 2(a) of main text where the pair level $(X, 1, 0) + (b, 2, 0)$ undergoes repulsion due to $(X, 2, 0) + (b, 1, 0)$. The field-dependency of $k_{\rm el}$ and $k_{2,\rm loss}$ for LiCs initially in $(X, 1, 0) + (b, 2, 0)$ are shown in Fig. 2(c) and (d). At $F \to 0$, due to shielding, $k_{2,\rm loss}$ is suppressed. However, at non-zero small fields, there is enhanced $k_{2,\rm loss}$ which is due to inelastic transitions to lower lying levels $(X, 1, m) + (b, 2, m')$, with $|m| + |m'| > 0$. The rotational levels are field dressed and hence the dipolar transitions $(\tilde{j}, 0) \leftrightarrow (\tilde{j}, \pm 1)$ are allowed. At very long-range ($R > 1000$ bohr), the $L = 2$ channels of such lower lying thresholds undergo avoided crossings with the incoming $L = 0$ channel of $(X, 1, 0) + (b, 2, 0)$, giving rise to the inelastic transitions. At higher fields, these avoided crossings occur at higher energies, which are energetically inaccessible at low collision energies. Hence, $k_{2,\rm loss}$ gets suppressed again.

The peaks in $k_{2,\rm loss}$ occur due to two reasons – one from FL states crossing the threshold, and the other due to the opening up of new inelastic channels as a function of field. It will also be interesting to study the dependency of $k_{2,\rm loss}$ on the collision energy and on the fine and hyperfine structure, which we have ignored in this study. A detailed study is, however, beyond the scope of the present work.

## S3. TRANSITION DIPOLE MOMENTS AND FRANCK-CONDON FACTORS

At large distances, properties of the tetramers can be expressed in terms of the properties of the dimers. Here we intend to calculate the transition dipole moment (TDM) and Franck-Condon factor (FCF) between states of $X + X$ and $X + b$ in the two-molecule basis set and express them in terms of single-molecule properties.

### A. Dipole moment matrix elements

The matrix element of a dipole operator $\hat{\boldsymbol{\mu}}_{\rm (AB)}$ for tetramer AB is expressed as the sum of individual dipole operators $\hat{\boldsymbol{\mu}}_{\rm (A)}$ and $\hat{\boldsymbol{\mu}}_{\rm (B)}$ defined for the molecules A and B

$$\hat{\boldsymbol{\mu}}_{\rm (AB)} \approx \hat{\boldsymbol{\mu}}_{\rm (A)} + \hat{\boldsymbol{\mu}}_{\rm (B)}. \tag{S8}$$

The dipole operators of the dimers $\hat{\boldsymbol{\mu}}_{\rm A(B)}$ act only on the individual dimer basis, and are diagonal in $L$ and $M_L$. So using Eq. (S1), we have

$$\langle \beta_{\rm A}, \tilde{j}_{\rm A}, m_{\rm A} | \hat{\boldsymbol{\mu}}_{\rm (A)} | \beta'_{\rm A}, \tilde{j}'_{\rm A}, m'_{\rm A} \rangle = \delta_{m_{\rm A} m'_{\rm A}} (-1)^{m_{\rm A}} \mu_{\beta_{\rm A} \beta'_{\rm A}}$$
$$\sum_{j_{\rm A} j'_{\rm A}} c^{\beta_{\rm A}, m_{\rm A}}_{j_{\rm A} \tilde{j}_{\rm A}} c^{\beta'_{\rm A}, m'_{\rm A}}_{j'_{\rm A} \tilde{j}'_{\rm A}} \sqrt{(2j_{\rm A}+1)(2j'_{\rm A}+1)} \begin{pmatrix} j_{\rm A} & 1 & j'_{\rm A} \\ 0 & 0 & 0 \end{pmatrix}$$
$$\times \begin{pmatrix} j_{\rm A} & 1 & j'_{\rm A} \\ -m_{\rm A} & 0 & m_{\rm A} \end{pmatrix}, \tag{S9}$$

and a similar expression for dimer B. For convenience we denote the RHS as $\mu_{\beta_{\rm A} \beta'_{\rm A}} d_{\tilde{j}_{\rm A} m_{\rm A}, \tilde{j}'_{\rm A} m'_{\rm A}}$, which separates the vibronic and the rotational parts.

### B. TDMs and FCFs between FL tetramers

Let $|\Psi_i\rangle$ and $|\Psi_f\rangle$ represent the wavefunctions for state $i$ in $X + X$ and state $f$ in $X + b$, respectively. We are interested in calculating the transition probabilities between these states, which are governed by their TDMs $D_{if}$ given by

$$D_{if} = \langle \Psi_i | \hat{\boldsymbol{\mu}}^{\rm T}_{\rm (AB)} | \Psi_f \rangle = \sum_{kk'} \Big( \int_0^\infty \psi^*_{i,k} \psi_{f,k'} dR \Big)$$
$$\times \langle \Phi_{i,k} | \hat{\boldsymbol{\mu}}^{\rm T}_{\rm (AB)} | \Phi_{f,k'} \rangle, \tag{S10}$$

where we have used the expansion of Eq. (S3). The superscript T denotes a TDM operator. The first factor in the RHS is the overlap of the radial wavefunctions, which we denote by $\mathcal{R}_{if,kk'}$. For bound wavefunctions, $\mathcal{R}_{if,kk'}$ is dimensionless, whereas for scattering states of $X + X$, $\mathcal{R}_{if,kk'}$ has dimensions of (energy)$^{-1/2}$ (if $|\Psi_i\rangle$ are energy normalized). The second factor is the expectation value of TDM in the coupled-channel basis. Using Eq. (S4) for the definition of the channel functions, we write the TDM matrix element as

$$\langle \Phi_{\rm g,i} | \hat{\boldsymbol{\mu}}^{\rm T}_{\rm (AB)} | \Phi_{\rm e,i'} \rangle = \langle X, \tilde{j}_{\rm A}, m_{\rm A}; X, \tilde{j}_{\rm B}, m_{\rm B}; L, M_L; + | \hat{\boldsymbol{\mu}}^{\rm T}_{\rm (AB)} | X, \tilde{j}'_{\rm A}, m'_{\rm A}; b, \tilde{j}'_{\rm B}, m'_{\rm B}; L', M'_L; + \rangle. \tag{S11}$$

Expanding the above matrix element in terms of the unsymmetrized functions, we obtain

$$\langle \Phi_{i,k} | \hat{\boldsymbol{\mu}}^{\rm T}_{\rm (AB)} | \Phi_{f,k'} \rangle = \frac{1}{2} \frac{\delta_{LL'} \delta_{M_L M'_L}}{\sqrt{(1+\delta_{AB})}}$$
$$\times \Big[ \mu^{\rm B}_{Xb} \Big( d_{\tilde{j}_{\rm B} m_{\rm B}, \tilde{j}'_{\rm B} m'_{\rm B}} \delta_{AA'} + d_{\tilde{j}_{\rm A} m_{\rm A}, \tilde{j}'_{\rm B} m'_{\rm B}} \delta_{BA'} \Big) + \mu^{\rm A}_{Xb} \Big( d_{\tilde{j}_{\rm A} m_{\rm A}, \tilde{j}'_{\rm B} m'_{\rm B}} \delta_{BA'} + d_{\tilde{j}_{\rm B} m_{\rm B}, \tilde{j}'_{\rm B} m'_{\rm B}} \delta_{AA'} \Big) \Big], \tag{S12}$$

where $\mu_{Xb}^{A(B)}$ is the vibrationally averaged transition dipole moment of dimer A(B). For identical molecules, Eq. (S12) reduces to

$$\langle \Phi_{i,k}|\hat{\boldsymbol{\mu}}_{(AB)}^{T}|\Phi_{f,k'}\rangle = \frac{\delta_{LL'}\delta_{M_L M_L'}}{\sqrt{(1+\delta_{AB})}}\mu_{Xb}\Big(d_{\tilde{j}_B m_B, \tilde{j}_B', m_B'}\delta_{AA'} + d_{\tilde{j}_A m_A, \tilde{j}_B', m_B'}\delta_{BA'}\Big), \tag{S13}$$

where italic $A(B)$ collectively represents the functions $|\beta_{A(B)}, j_{A(B)}, m_{A(B)}\rangle$. We denote the RHS by $\mu_{Xb}\mathcal{D}_{if,kk'}$, which is a product of an electronic transition dipole moment term and a dimensionless term representing the overlap between the rotational states of the different vibronic levels.

The final expression for the TDM matrix elements for the tetramers is thus obtained by collecting all the terms

$$D_{if} = \mu_{Xb}\sum_{kk'}\mathcal{R}_{if,kk'}\mathcal{D}_{if,kk'}. \tag{S14}$$

We denote the summation term in the above by $\mathcal{F}_{if}$, the square of which is a generalized FCF between the tetramer states that contains the overlap integral over all the internal coordinates of a tetramer. They are defined as

$$|\mathcal{F}_{if}|^2 = \Big|\sum_{kk'}\mathcal{R}_{if,kk'}\mathcal{D}_{if,kk'}\Big|^2. \tag{S15}$$

Both $\mathcal{R}_{if,kk'}(F)$ and $\mathcal{D}_{if,kk'}(F)$ are functions of the electric field. The potential wells in $X+X$ and $X+b$ change shapes with $F$, and so does $\mathcal{R}_{if,kk'}$. On the other hand, the dressed rotor functions are field dependent, which in turn imparts field dependency on $\mathcal{D}_{if,kk'}$. This makes the FCFs $|\mathcal{F}_{if}|^2$ tunable with $F$.

For a scattering state $|\Psi_i\rangle$ in $X+X$, the radial overlap matrix elements $\mathcal{R}_{if,kk'}$ have an energy dependency in addition to field dependency. This makes $|\mathcal{F}_{if}|^2$ as well as the $D_{if}$ energy dependent.


\fi
%--------------------------------------------------------------

\end{document}